\documentclass[11pt,twoside]{article}  
\usepackage{apn3conf}

\begin{document}   

\title{Radio characteristics of the very young Planetary Nebulae SAO\,244567 }

\author{Corrado Trigilio, Grazia Umana, Luciano Cerrigone\altaffilmark{1},
			}
          \affil{Istituto di Radioastronomia CNR,
		Sezione di Noto, Italy}
          \altaffiltext{1}{Istituto di Fisica e Astronomia, Universit\`a 
		di Catania, Italy}

\contact{Corrado Trigilio}
\email{c.trigilio@ira.cnr.it}

\paindex{Trigilio, C}
\aindex{Umana, G.}
\aindex{Cerrigone, L. }     

\authormark{Trigilio, Umana \&  Cerrigone }


\begin{abstract}          
The radio emission from the youngest known Planetary nebula, SAO\,244567,
has been mapped at 20,13,  6, 3.6 and 1.2 cm  by using the Australian 
Telescope Compact Array (ATCA). 
These observations constitute the first detailed radio study  of this very 
interesting object, as they allow us to obtain, for the first time, the 
radio morphology of the source and to compute the radio spectrum up 
to 18\,752 MHz.
\end{abstract}

\section{Introduction}

While the fate of a star with main sequence mass between 1 to 8 solar 
mass is well established, the formation and the early evolution of 
Planetary Nebulae (PN) is still one of the less understood phase of 
stellar evolution.

New clues on the process of PNs formation can be provided by the analysis 
of the physical characteristics of objects in the shortphase between 
the end of the AGB and the onset of the ionization in the nebula.
For this purpose, many authors have tried to identify very young PNs 
or proto Planetary Nebula (PPNs) but this is revealed to be quite 
difficult as this evolutionary phase is very rapid and because
the central object is often heavily obscured by the thick circumstellar
envelope formed during the AGB phase.

In this contest SAO\,244567 appears to be  a unique object:
\begin{itemize}
\item optical and ultraviolet spectra have changed in few decades 
(Parthasarathy et al., 1995);
\item it shows a strong infrared excess and is associated with 
IRAS\,17119-5926, with measured IRAS fluxes 0.65, 15.50, 8.20 and 3.52~Jy
at 12, 25, 60 and 100 $\mu$m respectively;
\item HST images of SAO\,244567 have revealed the presence of a structures 
1.6 arcsec nebula, where collimated outflows are evident 
(Bobrowsky et al., 1998).
\end{itemize}
These, together with other observational evidences make the source a 
perfect target for studying the early structure and evolution of PNs.

Despite of the numerous optical and ultraviolet studies of this very 
interesting object, very little is known on its radio properties.
Parthasarathy et al. (1993) briefly reported on ATCA  6 and 3 cm 
observations obtained in 1991. The measured flux densities are 
$63.3 \pm 1.8$ mJy and $51 \pm 12 $ mJy at 6 and 3 cm respectively,
the latest obtained by direct fitting of the UV data, yielding to a 
quite large error in the flux estimate.

SAO\,244567 is also associated to the radio source PMN\,J1716-5929, 
which is reported as a $43\pm 8$~mJy source at 4850 MHz (6~cm) 
in the Parkes-MIT-NRAO (PMN) survey.

\begin{figure}
\epsscale{1.0}
\plotone{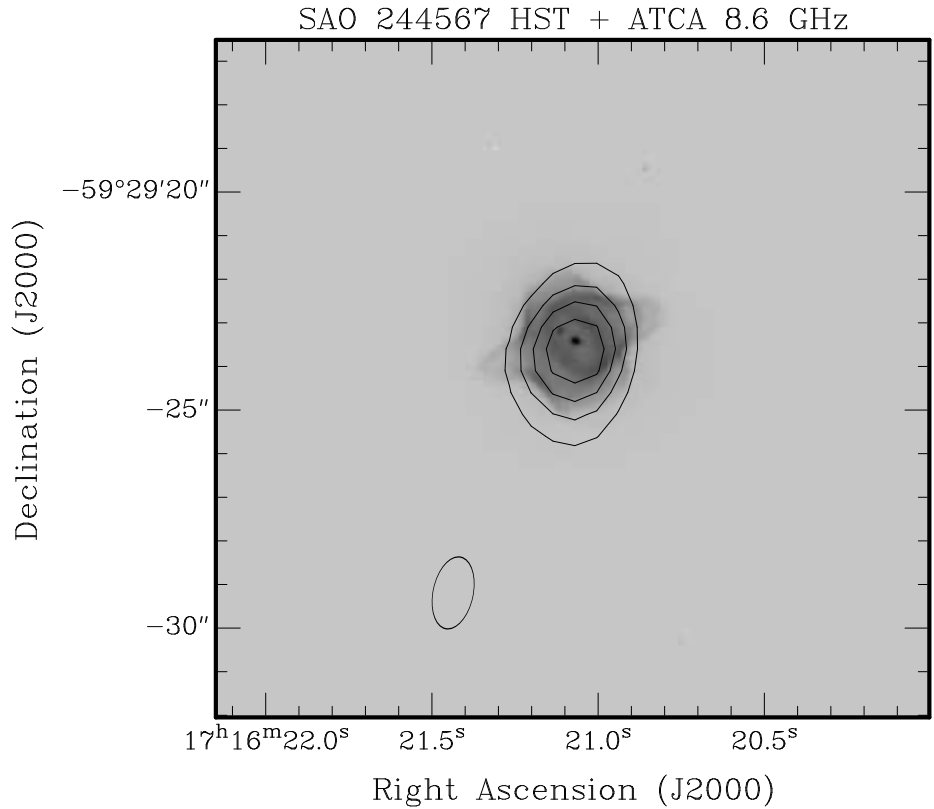}
\caption{The radio map of SAO\,244567 obtained with the ATCA at 8.6\,GHz
(contour levels) superimposed with the HST image in H$\alpha$
(from Bobrowsky et al., 1998). The synthetic radio beam is
shown in the lower left corner.} 
\label{map}
\end{figure}

\section{Observations and Results}
SAO\,244567 was observed with the Australia Telescope Compact Array 
(ATCA)  on August 24 and 29 2002. The array was in 6C configuration, 
having the highest baseline of 6 km, that, at 3.6cm, corresponds
to an angular resolution of $0.6^{\prime \prime}(\theta_\mathrm{FWHM}/3$).
On August 24 the observations were carried out simultaneously at 
16\,832 and 18\,752 MHz, by using the new millimetric receivers. 
At that time only 3 antennas were equipped with mm backends,
therefore the angular resolution at those frequencies was degraded.
On august 29 the observations were carried out simultaneously 
at 4800/8640 MHz and at 1384/2368. The flux scale was fixed toward 
1934-638, phase calibration performed by observing 1718-649.\\
~\\
The ATCA observations allow us to obtain for the first time the
radio morphology of the source and to compute the radio spectrum 
up to 18\,752\,MHz.
The observed flux densities are 36.8, 45.4, 47.9, 43.6, 36.1 and 36.0
respectively at 1384, 2368, 4800, 8640, 16\,832 and 18\,752 MHz. 
We note that the 4800 MHz flux density is about 25\% lower than what previously
observed by Parthasarathy et al. (1993) in 1991.

The 8\,640\,MHz radio map (Fig.~\ref{map}) reveals a quite extended structure, 
slightly elongated in the S-W direction (pa=$-12^o$), whose overall size
($1.68^{\prime\prime} \times 0.92^{\prime\prime}$) compares quite well with 
that observed in H$\alpha$ (Bobrowsky et al., 1998). The integrated
flux density is $43.6\pm 0.4$~mJy.

The computed radio spectrum is consistent with a free-free shell-like emitting 
nebula, whose angular size matches that estimated by our interferometric
measurements, with a decreasing density ($\propto r^{-2}$), a temperature 
$T\approx 10^4$K and $N_\mathrm{e}\approx 10^4\mathrm{cm^{-3}}$ in the innermost
part of the nebula. However, the ATCA angular
resolution prevented to point out any fine detail of the radio structure.

Once we have determined the angular size ($\theta_1\times\theta_2$) of
the source, we can use the measured radio flux density at 8.6\,GHz
($F_\mathrm{8.6\,GHz}$) to calculate some physical parameter that
characterize the nebula:\\
~\\
Brightness temperature
\begin{displaymath}
T_\mathrm{B}=25\frac{F_\mathrm{8.6\,GHz}}{\theta_1\times\theta_2}
=7\times 10^2\mathrm{K}
\end{displaymath}
Mean Emission Measure
\begin{displaymath}
EM=5.3\times 10^5 \frac{F_\mathrm{8.6\,GHz}}{\theta_1\times\theta_2}
=1.5\times 10^7\mathrm{cm^{-6}pc}
\end{displaymath}
Infrared Excess
\begin{displaymath}
IRE=\frac{FIR}{F_\mathrm{8.6\,GHz}}=2.87	
\end{displaymath}
where FIR is the total far infrared flux obtained by integrating 
a Planck curve fitting the IRAS fluxes.

\section{Conclusions}
A radio map at 8.6\,GHz (3.6 cm) of the very young Planetary Nebula
SAO\,244567 has been obtained, which reveals a slight extended radio
structure. The angular resolution of ATCA does not allow to evidence the 
fine details of the nebula as shown by HST observations.
However, the elongated radio structure give an hint of the presence of
two possible jets, in the S--W direction, that the synthetic beam 
($\theta_\mathrm{FWHM}=1.8^{\prime\prime}$) is not able to resolve.

The brightness temperature, the mean emission measure and the infrared
excess as derived by our observations are all consistent with a very 
young planetary nebula.

When compared with previous observations, the radio flux appears to vary.
In particular, the decrement of the 6~cm flux (from 1991 to 2002), while
the ionization of the nebula is increasing, does not agree with the recent 
models of the evolution of the radio luminosity from young Planetary Nebulae.

Further 5 frequencies radio monitoring and mapping with the ATCA fully
equipped with millimeter receivers are planned.
%
%
%
%


\end{document}